\documentclass[twocolumn,showkeywords,showpacs,preprintnumbers,amsmath,amssymb]{revtex4}
\usepackage{graphicx}% Include figure files
\usepackage{dcolumn}% Align table columns on decimal point
\usepackage{bm}% bold math

\begin{document}

\title{The P versus NP Problem in Quantum Physics}% Force line breaks with \\

\author{D. Song}

\affiliation{%
School of Liberal Arts, KoreaTech, Chungnam 330-708, Korea
}%

\date{\today}% It is always \today, today,

\begin{abstract}

Motivated by the fact that information is encoded and processed by physical systems, 
the P versus NP problem is examined in terms 
of physical processes.  In particular, we consider 
P as a class of deterministic,  and NP as nondeterministic, polynomial-time 
physical processes.  Based on these identifications, we review a self-reference physical 
process in quantum theory, which belongs to NP but cannot be contained in P.

\end{abstract}


\maketitle
\section{Introduction}
In the theory of computation, researchers have been thinking about 
impossible problems, difficult problems, and easy problems 
in the context of computational complexity \cite{complexity}. One of the major 
areas of interest has been a special type of problem \cite{cook1,cook2}, as a 
problem may be difficult to solve in general, but with the 
right choice, becomes easy to solve. There have been a 
lot of efforts in the computational science community to 
show whether these special types of hard problems, known 
as NP problems (or nondeterministic in polynomial time), 
are in fact easy ones, known as P problems (or deterministic in polynomial time), 
but the answer has remained elusive \cite{fortnow}.

As stated in the well-known phrase \lq\lq information is physical\rq\rq \cite{landauer0}, 
information processing is performed on a physical system. Therefore, 
computation should be related to the limits and possibilities of 
underlying laws of physics that govern the physical system \cite{landauer} (also see \cite{deutsch2}).  
Moreover, since the known laws of physics are quantum 
mechanics, the computing device should be a quantum 
Turing machine, rather than a classical Turing machine \cite{deutsch}. 
In the last few decades, people have indeed 
witnessed various aspects of quantum advantages that are unparalleled in 
classical computations. For instance, Bell's inequalities revealed one 
of the most profound quantum aspects unseen in the classical counterpart \cite{bell}. 
The quantum key distribution protocols \cite{bb84,ekert} have been shown to yield that 
quantumness can be strange and confusing, yet also beneficial to 
establishing secret keys. Moreover, various quantum algorithms \cite{deutsch,jozsa,shor,grover} 
have been shown to be computationally more efficient compared 
with previous algorithms that are run on classical computational systems. 
Therefore, it is natural to ask if such an advantage of quantum weirdness 
could shed light on other important problems including the P versus NP problem \cite{brassard,vazirani,tusarova,aaronson}.

\begin{figure}
\begin{center}
{\includegraphics[scale=.45]{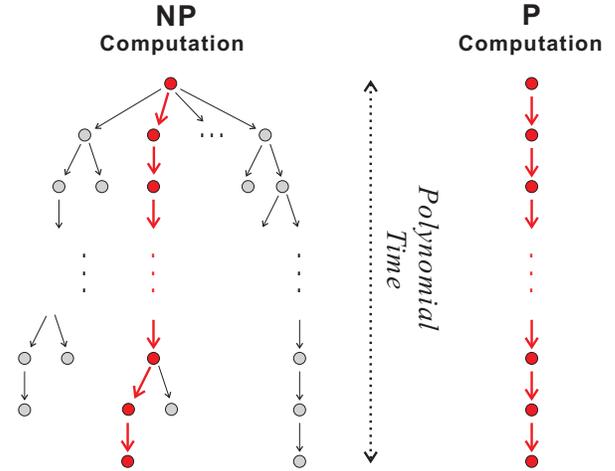}}

\end{center}
\caption{NP \& P Computations. The NP model has multiple ways to proceed the computation and 
chooses the path to reach the output in polynomial time. 
Unlike the P computation, commonly seen in ordinary computers,  
NP is a theoretical model of computation. }
\label{NPP}\end{figure}

NP computation is a theoretical model in which computing may proceed 
in nondeterministic ways, and chooses the acceptable path to reach the 
output in short time. It is noted that NP computation can 
also be defined indirectly as a difficult problem that can be 
verified in polynomial time in a physical computing device which is a verifier approach. However, 
we wish to consider NP computation in a direct way - a (nondeterministic) decider approach - i.e., this 
theoretical computing device indeed computes in a nondeterministic 
way and reaches the acceptable (i.e., by the computing entity) output in polynomial time, as shown in Fig. \ref{NPP}.
On the other hand, the case of deterministic computation in polynomial time, 
P computation, can be easily seen, since our ordinary computers follow this 
model of computation.

In this paper, we seek to denote NP as a class of nondeterministic   
 and P as a class of deterministic polynomial time physical processes. 
 That is, we will follow the logic of considering computation as a physical process.   
 In particular, we review a physical phenomenon of NP 
that has existed as a theoretical computing model and discuss this 
particular self-reference physical process cannot be computed by any computational 
devices (i.e., classical or quantum). 
It is noted that self-referencing has been one of the most effective tools in
proving a number of important theorems in logic.

This paper is organized as follows. In the second section, 
we review Turing machines, construct a particular nondeterministic 
computational model, and discuss deterministic computations 
that will be useful later. In the third section, we review a 
physical version of the computation and discuss the nondeterministic 
polynomial time physical process that cannot be computed by any deterministic computations. 
We conclude with brief remarks.

\section{Nondeterministic Computation}

\begin{figure}
\begin{center}
{\includegraphics[scale=0.9]{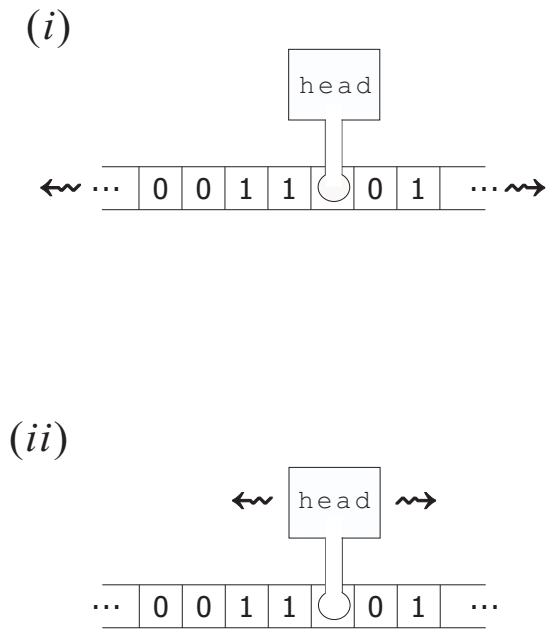}}

\end{center}
\caption{Symmetry in Turing Machines. It can be seen that whether the 
tape moves to the left or right as in (i), or the head moves to the right or left as in (ii), respectively, 
the readout should be the same.  In the case of quantum computation, 
whether the state (or the tape) evolves, 
the Schr\"odinger scheme, or the observable (or the head) moves  
to the opposite direction, the Heisenberg scheme, the outcome is the same.   }
\label{Turing}\end{figure}

In \cite{church,turing}, it is discussed that  
computation could be defined as something that may be performed 
on a certain type of machine, later to be known as Turing machines (TM). 
In general, TMs are defined with 
5-tuple $(Q,\Gamma,\delta,q_0,\Omega)$,  
where $Q$ corresponds to a set of states, 
$\Gamma$, an input alphabet, $\delta$, a transition function, $q_0$, an initial state, 
and $\Omega$, final states.  A nondeterministic TM (NTM) 
corresponds to a TM where given an input, the output may be more than one. 
Let us consider one particular NTM that will be useful later. 
Suppose in this particular NTM, for a given input, $a$, the output may be 
$b_1$ or $b_2$ where $b_1 \neq b_2$.  That is, 
\begin{equation}
\delta(a)=\{ b_1,b_2\}
\label{NTMtransition}\end{equation}
where $a\in \Gamma$ and $b_1,b_2 \in \Omega$ in 5-tuple of TM. 
In order to be more specific, we wish to define the transition function $\delta$ of NTM, i.e.  
$\delta\equiv \{ \delta_1, \delta_2\}$, such that, 
\begin{eqnarray}
\delta_1 &:& a\rightarrow b_1 \label{delta1} \\
\delta_2 &:& a\rightarrow b_2 \label{delta2}
\end{eqnarray}
As defined in NTM, $\delta_1$ and $\delta_2$ are applied 
nondeterministic way so that $a$ yields either 
$b_1$ or $b_2$, as defined in (\ref{NTMtransition}).

Deterministic Turing machines (DTM) may be considered with the notion of 
TM 5-tuple discussed above.  Unlike NTMs, the 
transition does not yield multiple processes.  Therefore, if we follow the 
NTM's definition of transition functions, we should consider that, 
in DTM, the computation proceeds equally under $\delta_1$ and $\delta_2$.  
This means that any DTM must have a transition rule where $\delta_1$ 
and $\delta_2$ yield the same computation, 
if we assume $\delta_1$ and $\delta_2$ are both valid transition functions.

It should be noted that NP is a theoretical concept of computation that is different from probabilistic computation. 
The NP computation is nondeterministic and may have multiple computational paths, given an input compared to the deterministic models. 
However, unlike the probabilistic model, which simply yields probability distribution 
for each output, the NP computation chooses the path for the output.  
In the next section, we will consider a physical example based on the NTM considered above, where the nondeterministic transitions 
correspond to the path choices, rather than probability distribution to the output.

\section{Quantum Case}
In quantum Turing machines, or quantum computation \cite{yao}, 
the classical states are replaced by quantum states and 
the transition functions correspond to unitary operations. 
In particular, there are two possible ways to perform computations. 
 The first way consists of applying unitary transition functions to the states, 
 while the second method involves applying the elementary gates to an observable. 
In both cases, the result would be the same.  
The first method is known as the Schr\"odinger scheme, while the latter is known as the Heisenberg 
scheme, named after the inventors of each method about a century ago.  
Quantum computation can be performed either by the first method \cite{deutsch,ekert2} 
or the second method \cite{heisenberg,hayden,vedral}. 
For example, an analogy can be made to the classical case, such that, when it is the tape 
that moves in the classical TM, it is like the Schr\"odinger picture, and when the head 
that reads the tape moves to the opposite direction, 
it corresponds to the Heisenberg picture (see Fig. \ref{Turing}). 
It can be easily seen that in both cases, the readout would be the same.

We now wish to consider one particular NTM that we defined previously in the quantum language. 
In the NTM in (\ref{NTMtransition}), we discussed that this 
particular machine has usual 5-tuple which can be made analogous 
with quantum computation.  However, we also wish to define the transition 
function $\delta$  
as follows: the $\delta_1$ transition function is identified with the Schr\"odinger 
operation and $\delta_2$ transition function is identified with the Heisenberg operation. 
These identifications are appropriate since both operations are valid as transition functions in quantum computation. 
On the other hand, DTM may now be defined with the case where the usual deterministic 
transition operations including $\delta_1$ and $\delta_2$ yield 
the same computational process.

To consider physical processes corresponding to NTM and DTM discussed above, 
we now briefly outline the result obtained in \cite{song}.
Let us assume the qubit, a unit vector 
in Bloch sphere notation, is initially pointing at
$z$-direction, ${\hat{\mu}}=(0,0,1)$, and also note the
corresponding observable as ${\hat{\nu}} =(0,0,1)$.   
A unitary transformation 
on the system vector ${\hat{\mu}}$ is then considered with a rotation about $y$-axis by
$\theta$, that is, $U_{\theta}$.  The unitary operation $U_{\theta}$
transforms the input as $ U_{\theta} {\hat{\mu}} U_{\theta}^{\dagger}$, 
that is, the operation under the transition function $\delta_1$.  
With $\delta_2$, $U_{\theta}^{\dagger}$ transforms the vector representing
the observable into $U_{\theta}^{\dagger}{\hat{\nu}} U_{\theta}$. 
It can be checked that the expectation value of
$(U_{\theta}^{\dagger} {\hat{\nu}} U_{\theta})\cdot {\hat{\mu}}$
for $\delta_2$ is equal to the expectation value under $\delta_1$, that is, ${\hat{\nu}} \cdot (U_{\theta} {\hat{\mu}}
U_{\theta}^{\dagger})$. Therefore, in the first case, both transition functions yield identical results.

We now wish to employ the technique of self-referencing; that is,
when the input state to be transformed is the coordinate vector itself,
i.e., ${\hat{\mu}}={\hat{\nu}}$. Since the coordinate vector is
also a vector, it is certainly possible for the vector
to be a system vector to be transformed. With $\delta_1$, the evolution is as follows:
\begin{equation}
\delta_1 : {\hat{\nu}} \rightarrow {\hat{\nu}}^{\prime} =
(\sin\theta,0,\cos\theta)
\label{path1}\end{equation}
Moreover, quantum theory provides a second approach where it
is the coordinate vector that is rotated counter-clockwise.  Therefore,
the same vector is transformed as
\begin{equation}
\delta_2 : {\hat{\nu}} \rightarrow {\hat{\nu}}^{\prime\prime} =
(-\sin\theta,0,\cos\theta)
\label{path2}\end{equation}
It is noted that
${\hat{\nu}}^{\prime} \neq {\hat{\nu}}^{\prime\prime}$ unless
$\theta = k\pi$ where $k$ is an integer.
We may identify ${\hat{\nu}}$ with the initial 
state $a$ of NTM and $b_1$ and $b_2$ in (\ref{delta1}) and (\ref{delta2})  
with ${\hat{\nu}}^{\prime}$ and ${\hat{\nu}}^{\prime\prime}$ then the above case corresponds 
to a nondeterministic computation as in (\ref{NTMtransition}).

Let us discuss to what physical processes the above description
corresponds.  
While the first case, i.e., where the input state is ${\hat{\mu}}$, is a common description of a single qubit operation, the second case
is a special case of the first one, i.e., when the input is the vector
representing a coordinate, ${\hat{\mu}}={\hat{\nu}}$. 
The physical phenomena description of these two cases can be stated as follows: 
\\
\\
{\bf{(P1)}} {\it{An observer observes the unitary evolution of ${\hat{\mu}}$ with respect to ${\hat{\nu}}$}}.
\\
\\
{\bf{(P2)}} {\it{An observer observes the unitary evolution of ${\hat{\nu}}$ with respect to ${\hat{\nu}}$}}.
\\
\\
It is noted that the physical process {\bf{(P1)}} is perfectly and sufficiently 
computed with a quantum computer, with input ${\hat{\mu}}$ and the
coordinate vector, or reference frame, ${\hat{\nu}}$ as we described above where $\delta_1$ and $\delta_2$ 
yield the same outcome in a deterministic way. However, is {\bf{(P2)}} a physical process, as well? 
It is certainly possible for the observer to choose ${\hat{\nu}}^{\prime}$ or ${\hat{\nu}}^{\prime\prime}$ 
when there is no qubit to measure. 
This is a unique observation because the experience of choosing 
reference frames without surroundings is not seen in classical
physics. This peculiar self-reference physical phenomenon corresponds to
{\bf{(P2)}}.

Therefore, the physical process of {\bf{(P2)}} 
corresponds to a nondeterministic computation as in (\ref{NTMtransition}). 
In particular, it is noted that the transition of 
${\hat{\nu}}$ into ${\hat{\nu}}^{\prime}$ and ${\hat{\nu}}^{\prime\prime}$ 
under two transition operations, i.e., (\ref{path1}) and (\ref{path2}), respectively, corresponds to the decisions or 
choices being made by the observer in {\bf{(P2)}}.  
Moreover, since the operations from the initial into final states is just 
a single rotation operation, this physical process 
certainly belongs to the category of the NP.  
 As discussed, any DTM should involve 
both $\delta_1$ and $\delta_2$ as valid transition operations.  In the 
case of quantum computation, deterministic computation should be equivalent under 
both the Schr\"odinger and the Heisenberg pictures.   
The single-operation nondeterministic physical process {\bf{(P2)}} cannot 
be computed by any quantum computer that always yields the same outcome under both schemes \cite{song}.  
It is noted that DTM cannot compute either path of (\ref{path1}) or (\ref{path2}).

Can classical TM compute the NP proposed above?  The answer is no.  
The nondeterministic physical process discussed above is a quantum phenomenon that cannot 
be observed in the classical computational models.  
It is clear that in the classical TM, as shown in Fig. \ref{Turing}, there is no classical 
physical process, i.e., where the head and the tape movements yield two different results.  
Therefore, {\bf{(P2)}} belongs to NP, a class of nondeterministic polynomial-time physical processes, but not in P, a class of 
deterministic polynomial time physical processes.

As previously noted, NP is also considered, indirectly, to be a polynomial time 
verifiable computation rather than the nondeterministic decider 
we considered above.  
Based on this verifier description of NP, it seems that 
the physical process described above can be verified by 
just two computational steps of DTM;  however, this does not work. 
For instance, if we consider NP in {\bf{(P2)}} indirectly, 
and based on our identification of computation as a physical process,   
 while it is difficult to know which choice is the physical path, but it is easy to verify that 
 the path is indeed the physical process when given the choice.  
However, as discussed above, there are not any deterministic machines that can compute paths (\ref{path1}) 
or (\ref{path2}); therefore, the indirect NP 
argument of hard to solve but easy to verify does not work in this case.

\section{Remarks}

In this paper, we have reviewed a  
physical phenomenon {\bf{(P2)}} that corresponds to NP computation. 
That is, when the reference frame itself is also being observed, this unique phenomenon corresponds to the case of 
NP computation where two valid transition functions yield 
two different path choices made by the observer.  As argued, this unique 
physical version of NP is not present in ordinary physical processes - in this case,  
the deterministic computational model where the 
two transition functions always yield the same outcome.

It is also interesting to note that even the extremely simple physical process of nondeterministic polynomial 
time in quantum physics, cannot be computed by any deterministic computations. 
Similar to the quantum algorithms, and 
quantum key distributions, quantum weirdness again yields 
another advantageous effect that is unseen in 
the classical computation system in the case of the P versus NP problem.

%%%%%%%%%%%%%%%%%%%%%%%%%%%%%%%%%%%%%%%%%%%%%%%%%%%%%%%%%%%%%%%%%%%%%%%%
%%%%%%%%%%%%%%%%%%%%%%%%%%%%%%%%%%%%%%%%%%%%%%%%%%%%%%%%%%%%%%%%%%%%%%%%
%%%%%%%%%%%%%%%%%%%%%%%%%%%%%%%%%%%%%%%%%%%%%%%%%%%%%%%%%%%%%%%%%%%%%%%%
%%%%%%%%%%%%%%%%%%%%%%%%%%%%%%%%%%%%%%%%%%%%%%%%%%%%%%%%%%%%%%%%%%%%%%%%
%%%%%%%%%%%%%%%%%%%%%%%%%%%%%%%%%%%%%%%%%%%%%%%%%%%%%%%%%%%%%%%%%%%%%%%%


\begin{thebibliography}{}
\bibitem{complexity} O. Goldreich, Computational complexity: a conceptual perspective, Cambridge University press (2008).

\bibitem{cook1} S. Cook, The complexity of theorem proving procedures, Proceedings Third Annual ACM Symposium on Theory of Computing, 151 (1971).

\bibitem{cook2} S. Cook, The P versus NP problem, \newline  http://www.claymath.org/sites/default/files/pvsnp.pdf.

\bibitem{fortnow} L. Fortnow, The status of the P versus NP problem, Comm. ACM, {\bf{52}}, 78 (2009). 



\bibitem{landauer0} R. Landauer, Information is physical, Phys. Tod. May, 23 (1991).

\bibitem{landauer} R. Landauer, Physical nature of information, Phy. Lett. A {\bf{217}}, 188 (1996).

\bibitem{deutsch2} D. Deutsch, A. Ekert, and R. Lupacchini, Machines, logic and quantum physics, Bull. Symbolic Logic, {\bf{6}} (3), 265 (2000); arXiv:math/9911150 [math.HO].

\bibitem{deutsch} D. Deutsch, Quantum theory, the Church-Turing principle and the universal quantum computer, Proc. R. Soc. London A {\bf{400}}, 97
(1985).


\bibitem{bell} J.S. Bell, On the Einstein-Podolsky-Rosen paradox. Phys. {\bf{1}}, 195 (1964).

\bibitem{bb84} C.H. Bennett and G. Brassard, Quantum cryptography: public key distribution and coin tossing, 
Proceedings of IEEE International Conference on Computers, Systems and Signal Processing, Bangalore, India, December, 175 (1984).

\bibitem{ekert} A.K. Ekert, Quantum cryptography based on Bell's theorem. Phys. Rev. Lett. {\bf{67}}, 661 (1991).



\bibitem{jozsa} D. Deutsch and R. Jozsa,  Rapid solution of problems by quantum computation. Proc. R. Soc. Lond. A {\bf{439}}, 553 (1992).

\bibitem{shor} P. Shor, Algorithms for quantum computation: discrete logarithms and factoring, in Proceedings of the 35th Annual Symposium on the Foundations of Computer Science, edited by S. Goldwasser
(IEEE Computer Society, Los Alamitos, CA), P. 124.; arXiv:quant-ph/9508027.

\bibitem{grover} L.K. Grover, Quantum mechanics helps in searching for a needle in a haystack, 
Phys. Rev. Lett. {\bf{79}}, 325 (1997); arXiv:quant-ph/9706033.

\bibitem{brassard} C.H. Bennett, E. Bernstein, G. Brassard, and U. Vazirani, 
Strengths and weaknesses of quantum computing, SIAM J. Comp. {\bf{26}}, 1510 (1997); arXiv:quant-ph/9701001. 

\bibitem{vazirani} U. Vazirani, A survey of quantum complexity theory, Proc. Sympos. Appl. Math. {\bf{58}} (2002).

\bibitem{tusarova} T. Tusarova, Quantum complexity classes, \newline arXiv:cs/0409051 [cs.CC].

\bibitem{aaronson} S. Aaronson, NP-complete problems and physical reality, ACM SIGACT
News Complexity Theory Column, March. ECCC TR05-026 (2005); arXiv:quant-ph/0502072.

\bibitem{church} A. Church, An unsolvable problem in elementary number theory, Am. J. Math. {\bf{58}}, 345 (1936).

\bibitem{turing} A.M. Turing, On computable numbers, with an application to the Entscheidungsproblem, Proc. London Math. Soc. (2) {\bf{442}}, 230
(1936).





\bibitem{yao} A. Yao, Quantum circuit complexity, Proceedings 34th Annual Symposium on Foundations of Computer Science (FOCS), 352 (1993).

\bibitem{ekert2} A.K. Ekert and R. Jozsa, Quantum computation and Shor's factoring algorithm. Rev.  Mod. Phys. {\bf{68}}, 733 (1996).

\bibitem{heisenberg} D. Gottesman, The Heisenberg representation of quantum computers, arXiv:quant-ph/9807006v1.

\bibitem{hayden} D. Deutsch and P. Hayden, Information flow in entangled quantum systems,  Proc. R. Soc. Lond. A {\bf{456}}, 1759 (1999); arXiv:quant-ph/9906007.

\bibitem{vedral} C. Hewitt-Horsman and V. Vedral, Developing the Deutsch-Hayden approach to quantum mechanics, New J. Phys. {\bf{9}}, 135 (2007); arXiv:quant-ph/0609085.

\bibitem{song} D. Song, Non-computability of consciousness, NeuroQuant. {\bf{5}}, 382 (2007); arXiv:0705.1617 [quant-ph].



\end{thebibliography}
\end{document}